\documentclass[seceq,jjapprn]{jjap}
\usepackage{graphicx}
\usepackage{lineno}

\title{First-Principles Study on Structural Properties of GeO$_2$ and SiO$_2$ under Compression and Expansion Pressure}
\author{Shoichiro~\textsc{Saito}\footnote{E-mail address: saito@cp.prec.eng.osaka-u.ac.jp} and Tomoya~\textsc{Ono}}
\inst{Graduate School of Engineering, Osaka University, 2-1 Yamadaoka, Suita, Osaka 565-0871, Japan }
\linenumbers
\abst{The detailed analysis of the structural variations of three GeO$_2$ and SiO$_2$ polymorphs ($\alpha$-quartz, $\alpha$-cristobalite, and rutile) under compression and expansion pressure is reported. First-principles total-energy calculations reveal that the rutile structure is the most stable phase among the phases of GeO$_2$, while SiO$_2$ preferentially forms quartz. GeO$_4$ tetrahedras of quartz and cristobalite GeO$_2$ phases at the equilibrium volume are more significantly distorted than those of SiO$_2$. Moreover, in the case of quartz GeO$_2$ and cristobalite GeO$_2$, all O-Ge-O bond angles vary when the volume of the GeO$_2$ bulk changes from the equilibrium point, which causes further deformation of tetrahedra. In contrast, the tilt angle formed by Si-O-Si in SiO$_2$ markedly changes. This flexibility of the O-Ge-O bonds reduces the stress at the Ge/GeO$_2$ interface due to the lattice-constant mismatch and results in the low defective interface observed in the experiments [Matsubara \textit{et al.}: Appl. Phys. Lett. \textbf{93} (2008) 032104; Hosoi \textit{et al.}:  Appl. Phys. Lett. \textbf{94} (2009) 202112].}
\kword{}
\begin{document}
\maketitle
\sloppy
\section{Introduction}
Ge has recently attracted increasing attention for future advanced complementary metal oxide semiconductor (CMOS) device structures owing to its high intrinsic carrier mobility as it is becoming increasingly difficult to enhance the performance of CMOS devices through scaling based on conventional Si-based techniques. The key issue to be resolved in advanced Ge-based devices is the formation of gate stacks with superior interface properties. Research activity related to GeO$_2$ is steadily increasing, and it has been reported that Ge is well passivated by GeO$_2$ by conventional dry oxidation without any hydrogen passivation treatment \cite{takagi,hosoi}. On the theoretical side, Houssa \textit{et al}. claimed that the viscoelastic properties of GeO$_2$ lead to a low interface defect density at the Ge/GeO$_2$ interface after performing a calculation using a modified Maxwell model \cite{houssa}. In our previous study, we examined the oxidation mechanism of crystalline GeO$_2$ and Ge(100) interfaces by a first-principles total energy calculation following the Si(100) oxidation process proposed by Kageshima and Shiraishi \cite{kageshima}, and we found that Ge atom emission, which deteriorates the Ge/GeO$_2$ interface, hardly occurs during the oxidation process of Ge(100) \cite{saito}. To investigate the low probability of Ge atom emission, the mechanism accounting for the release of interface stress should be clarified. 

From the viewpoint of phase transition, SiO$_2$ occurs in many different forms. At ambient temperature and pressure, the ground-state structure for SiO$_2$ is $\alpha$-quartz (q-SiO$_2$). SiO$_2$ forms a rutile structure (r-SiO$_2$) under a pressure above 2 GPa and it is transformed into $\alpha$-cristobalite (c-SiO$_2$) at a high temperature. On the other hand, there are two stable polymorphs of GeO$_2$ at normal pressures: the low temperature form has the rutile structure (r-GeO$_2$), and GeO$_2$ undergoes a smooth transformation to $\alpha$-quartz (q-GeO$_2$) at $T\approx$ 1300 K. An $\alpha$-cristobalite structure (c-GeO$_2$) has been identified after the long-time heating of GeO$_2$ glass at 873 K \cite{boehm}. The investigation of the c-GeO$_2$ phase is important because Ge substrates are typically subjected to dry oxidation in an O$_2$ ambient at 623-823 K to form a GeO$_2$ layer in recent experiments \cite{hosoi,temp}. 

Here, we investigate the structural properties of q-, c-, and r-GeO$_2$ by first-principles total energy calculations. The structural properties of SiO$_2$ are also examined for comparison. It was found that the rutile structure of GeO$_2$ is the most stable structure, whereas SiO$_2$ preferentially forms the quartz structure. We then investigate the variations of the O-Ge-O (O-Si-O) bond angles of quartz and cristobalite phases with respect to the volume since local pressure is induced at the semiconductor/oxide interface. Although the variations of the atomic structures of q-GeO$_2$, r-GeO$_2$, q-SiO$_2$, c-SiO$_2$, and r-SiO$_2$ under pressure have been examined by both experimental and theoretical studies \cite{33,25,32,c1,c2,1992,1994,1996,demuth,2000-1,chelikowsky,2005,h}, no study reports the relationship between their bond angles and pressure of these six oxides in the same treatment of computational code or experimental facility. We note that the lattice constant (bulk modulus) of q-SiO$_2$ reported by the first-principles calculation \cite{1996} is smaller (larger) by more than 3 $\%$ (34 $\%$) than the one reported by another first-principles study \cite{demuth}. Moreover, it is reported that a dioxide forms a cristobalite structure before the atom emission as well as at the initial stage of oxidation \cite{kageshima}. The pressure-dependent behavior of c-GeO$_2$, which has never been explored to the best of our knowledge, is a subject of intense research to clarify the relaxation mechanism of the interface stress. Therefore, it is of importance that the uniform theoretical treatment facilitates systematic comparisons and the identification of trends among these six oxides. Our finding is that the O-Ge-O bond angles change significantly under pressure, while the tilt angle and Si-O bond length vary in the case of SiO$_2$. The variation of the bonding network of GeO$_2$ exhibits completely different characteristics from that of SiO$_2$. The metallic properties of Ge provide a qualitative understanding of not only the difference between the ground-state phases of GeO$_2$ and SiO$_2$ but also the variations of the bond angles under pressure. 

The rest of this paper is organized as follows. In \S~2, we describe the computational techniques used in this study. In \S~3, we present the main results and a discussion of the structural parameters and properties. Finally, a brief summary is given in \S~4. 

\section{Computational Techniques}
\begin{figure}
\includegraphics{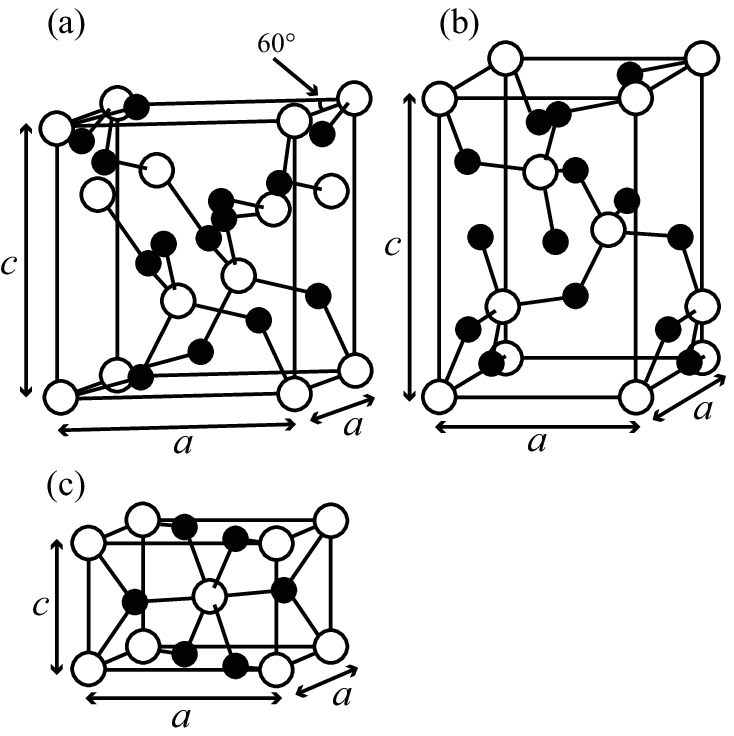}
\caption{Unit cells of quartz (a), cristobalite (b), and rutile (c). Black and white circles are O and Ge (Si) atoms, respectively. }
\label{fig:unit}
\end{figure}

The structures of q-, c-, and r-GeO$_2$ are hexagonal, tetragonal, and tetragonal with three, four, and two GeO$_2$ molecules per unit cell, respectively. Ge atoms in q- and c-GeO$_2$ are surrounded by four oxygen atoms, while each Ge atom in r-GeO$_2$ is surrounded by six oxygen atoms with distorted octahedral coordination as shown in Fig.~\ref{fig:unit}. The calculations are performed within the local density approximation \cite{PZ} of density functional theory \cite{HK,KS} using the real-space finite-difference approach \cite{Che,Che2,hirose,ono,ono2} and the norm-conserving pseudopotentials \cite{kobayashi} of Troullier and Martins \cite{TM} in the Kleinman-Bylander representation \cite{KB}. The grid spacing was set at 0.25 bohr, and a denser grid spacing of 0.083 bohr in the vicinity of nuclei with the augmentation of double-grid points \cite{ono} for each GeO$_2$ polymorph. We took $4 \times 4 \times 4$, $4 \times 4 \times 3$, and $4 \times 4 \times 6$ \textit{k}-point grids in the Brillouin zone for q-, c-, and r-GeO$_2$, respectively. The optimal lattice parameters and internal atomic coordinates were determined by minimization of the total energy using calculated forces, with a force tolerance of F$_{max} <$ 1.0 mH/bohr. The same computational procedures were applied for SiO$_2$ polymorphs.

\section{Results and Discussion}
\begin{figure}
\includegraphics{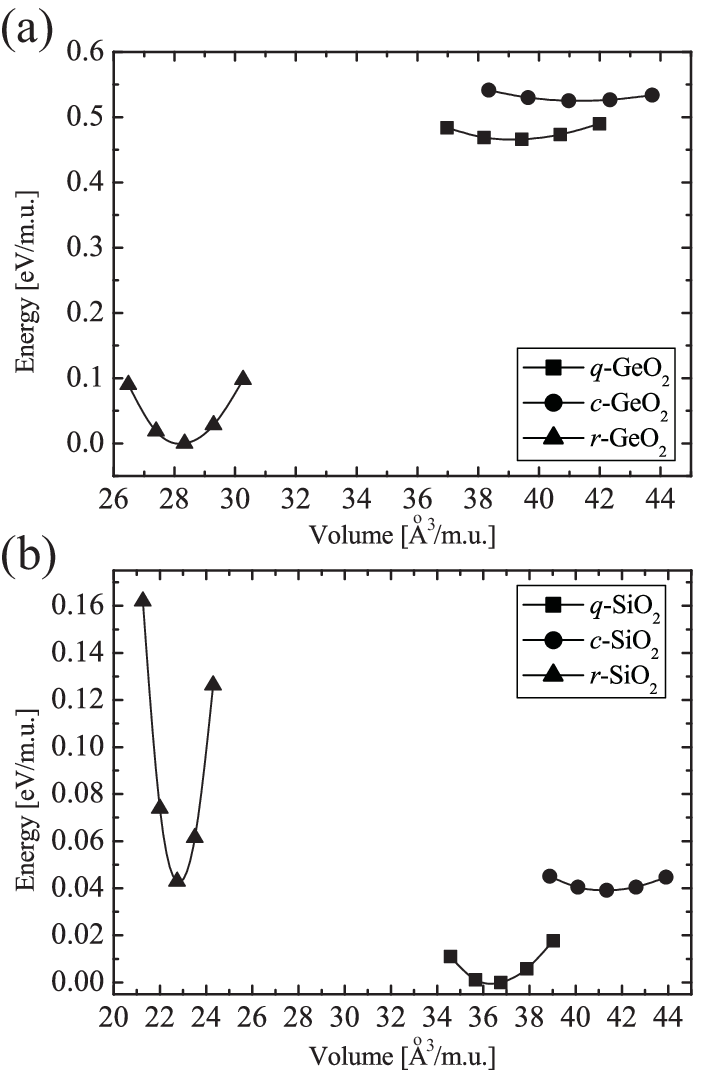}
\caption{Total energy per molecular unit (m.u.) as a function of volume for all polymorphs of GeO$_2$ (a) and SiO$_2$ (b). The zero of the energy scale is rutile for GeO$_2$ and quartz for SiO$_2$.}
\label{fig:energy}
\end{figure}

Figures~\ref{fig:energy} (a) and (b) show the total energy per molecular unit as a function of volume for the three structures of GeO$_2$ and SiO$_2$, respectively. The zeros of the energy scales are rutile for GeO$_2$ and quartz for SiO$_2$. It was found that the zero-temperature phase in GeO$_2$ has the rutile structure while that in SiO$_2$ exhibits the quartz structure, which is in good agreement with a previous report \cite{chelikowsky,comm}. Sn, which is also a group IV element similarly to Si and Ge, is a metal and its oxide crystallizes in the rutile structure under ambient conditions \cite{sno2}. GeO$_2$ forms the sixfold-coordinated rutile structure more preferentially than SiO$_2$ because Ge is between Si and Sn in the periodic table. Tables~\ref{tbl:lattice} and \ref{tbl:angle} show the calculated lattice constants, bond lengths $l_i$, bond angles $\theta_j$, and tilt angles $\delta$ of GeO$_2$ and SiO$_2$. The other calculated and experimental results are also given \cite{h,smith,hazen,25,pluth,51,32}. There are two distinct Ge-O (Si-O) bond lengths in GeO$_4$ (SiO$_4$) tetrahedra. In addition, $\theta_j$ exactly corresponds to the O-Ge-O (O-Si-O) bond angle, and the tilt angle is related to the Ge-O-Ge (Si-O-Si) bond angle \cite{tilq, tilc}. The agreement between our results and experimental results for SiO$_2$ is excellent for the structural parameters. The lattice constants of q- and c-GeO$_2$ are slightly underestimated in both the theoretical calculations: this underestimation is caused by the use of the local density approximation \cite{PZ}, and the parameters obtained by the theoretical calculations agree well. The deviations of the O-Ge-O bond angles from the ideal tetrahedral angle (109.5$^{\circ}$) are larger than those for O-Si-O, resulting in distorted GeO$_4$ tetrahedra. 

\begin{figure}
\includegraphics{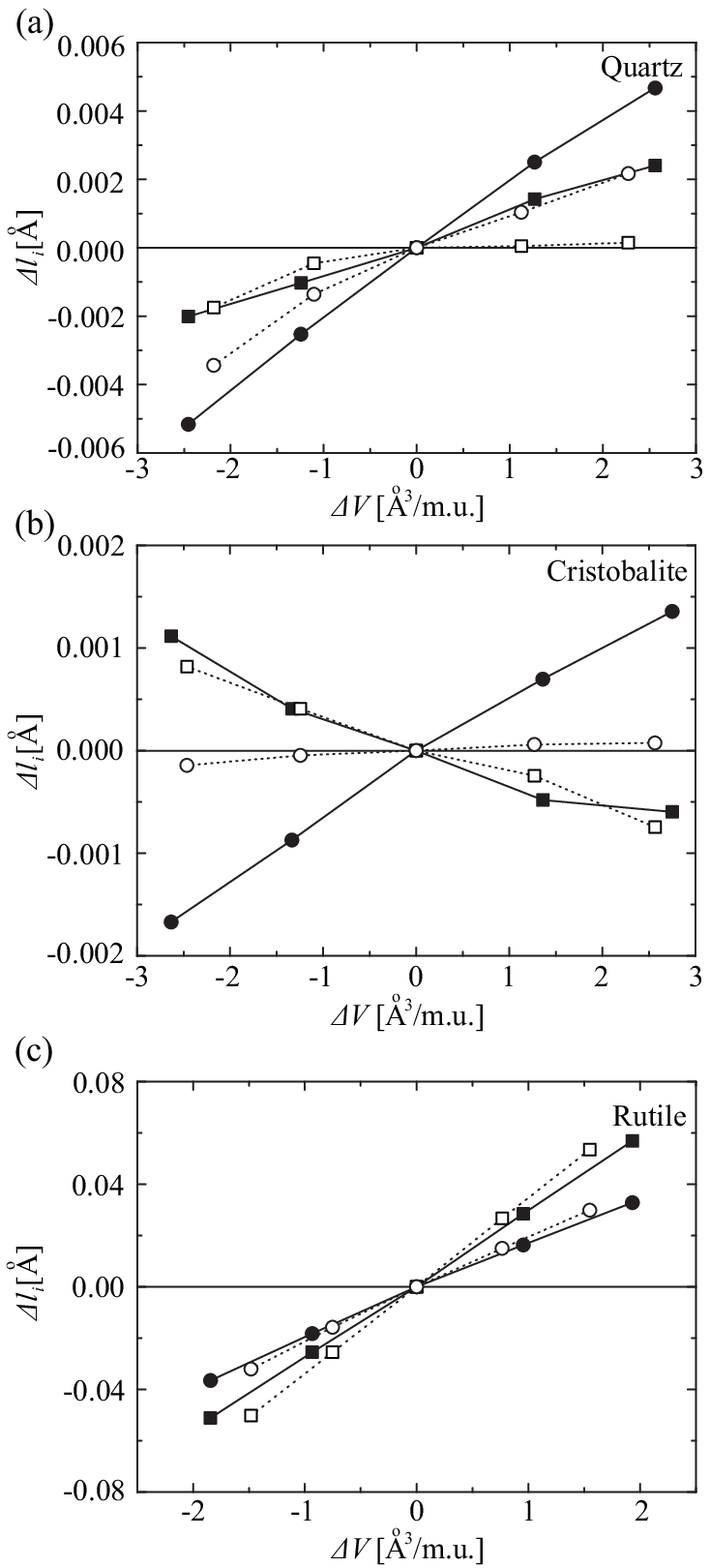}
\caption{Variations of x-O bond lengths $\Delta l_i$ in quartz (a), cristobalite (b), and rutile (c) structures from their equilibrium points. x represents a Ge or Si atom. $\Delta l_1$ and $\Delta l_2$ correspond to squares and circles, respectively. Black (white) symbols are the results of GeO$_2$ (SiO$_2$) and lines are only eye guides.}
\label{fig:length}
\end{figure}

\begin{figure}
\includegraphics{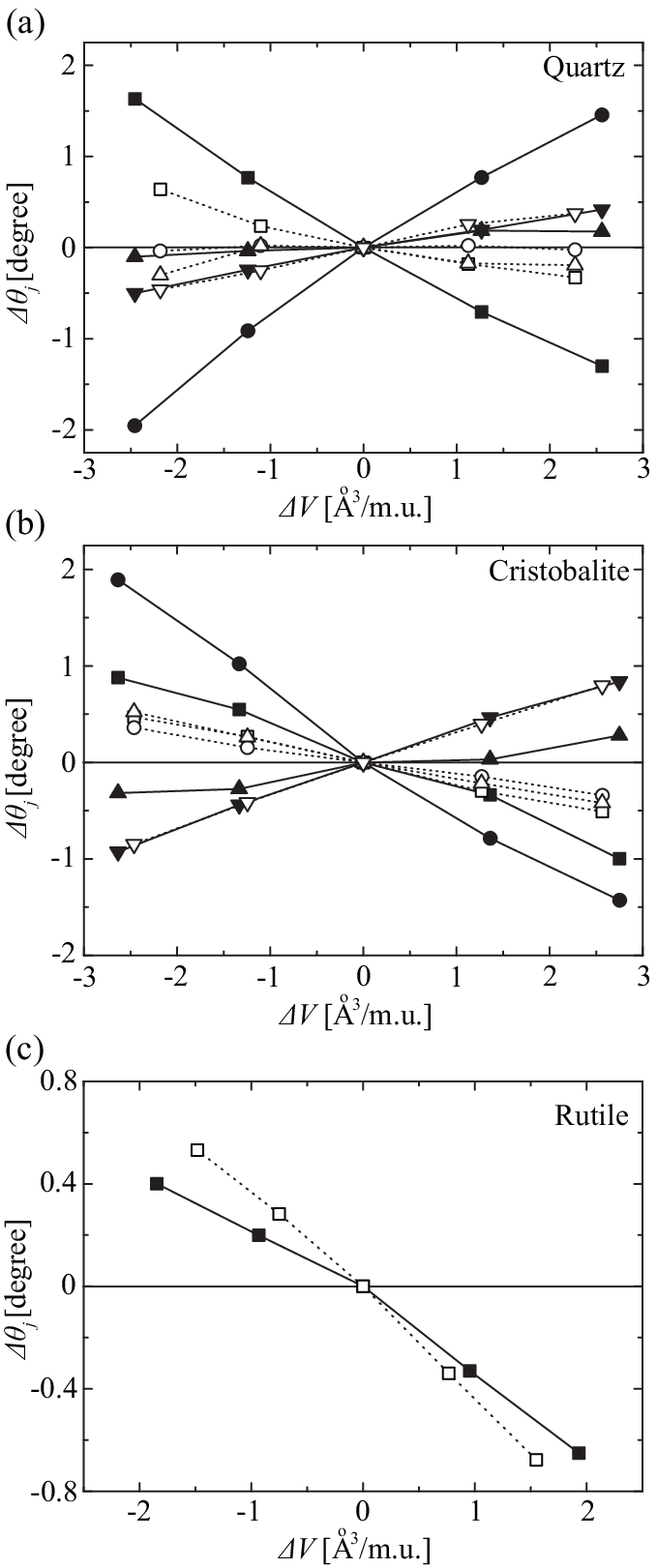}
\caption{Variations of O-x-O bond angles $\Delta \theta_j$ in quartz (a), cristobalite (b), and rutile (c) structures from their equilibrium points. x represents a Ge or Si atom. $\Delta \theta_1$, $\Delta \theta_2$, $\Delta \theta_3$, and $\Delta \theta_4$ correspond to squares, circles, upper triangles, and lower triangles, respectively. Black (white) symbols are the results of GeO$_2$ (SiO$_2$) and lines are only eye guides.}
\label{fig:O_x_O}
\end{figure}

\begin{figure}
\includegraphics{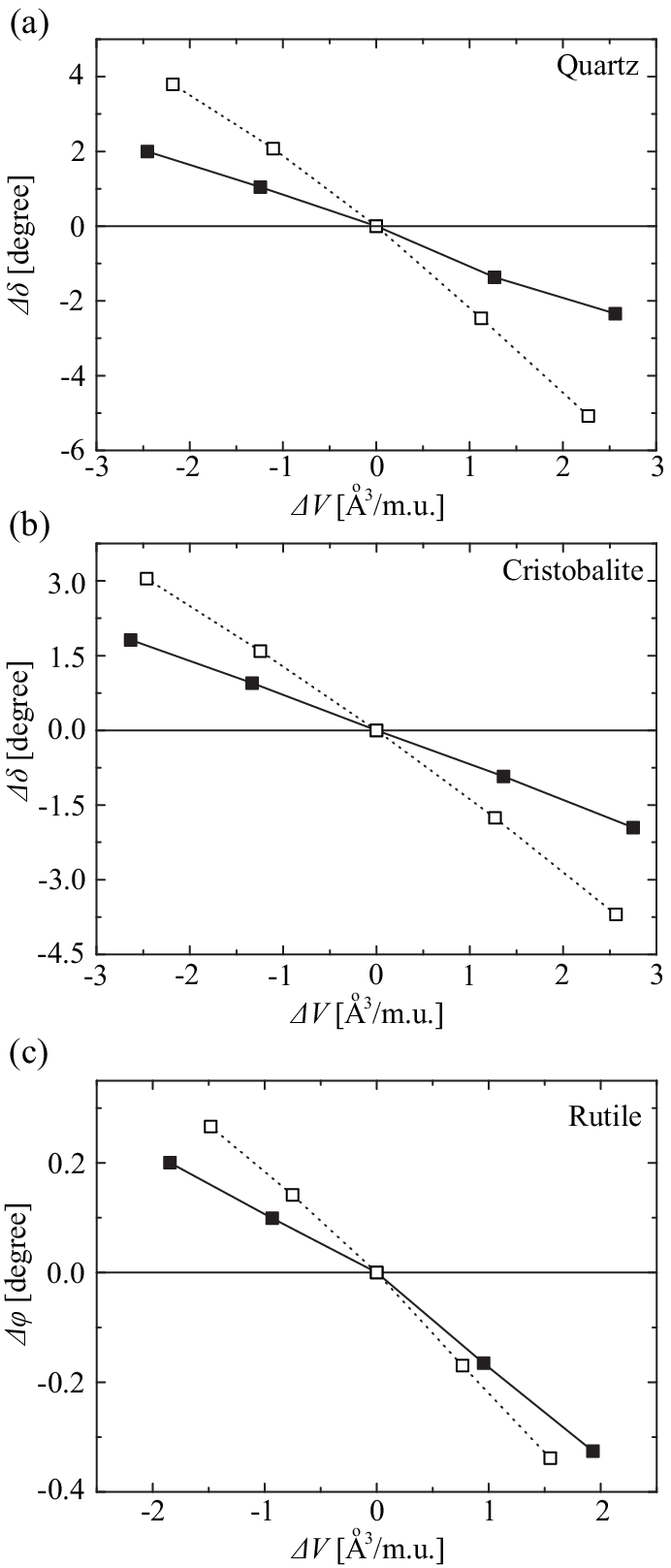}
\caption{Variations of tilt angles $\Delta \delta$ in quartz (a) and cristobalite (b) structures from their equilibrium points. $\Delta \delta$ is related to x-O-x bond angles. Variations of x-O-x bond angles $\Delta \varphi$ in rutile phases from their equilibrium points are shown in (c). x represents a Ge or Si atom. Black (white) symbols are the results of GeO$_2$ (SiO$_2$) and lines are only eye guides.}
\label{fig:x_O_x}
\end{figure}

Thermally oxidized GeO$_2$ on a Ge substrate has been found to mainly form the fourfold-coordinated structure by X-ray photoelectron spectroscopy \cite{xps}, and crystalline q- and c-SiO$_2$ are formed on a Si(100) surface according to the oxidation model of the Si(100) surface \cite{kageshima}. The compressive in-plane stress at the Ge/GeO$_2$ (Si/SiO$_2$) interface is induced by the lattice mismatch between Ge (Si) and its oxide. Therefore, we particularly focus on the bond structures of crystalline q- and c-GeO$_2$ assuming that the oxidation mechanism of Ge is the same as that of Si. Figures~\ref{fig:length}, ~\ref{fig:O_x_O}, and ~\ref{fig:x_O_x}  show the variations of the Ge-O (Si-O) bond lengths, the O-Ge-O (O-Si-O) bond angles, and the tilt angles with respect to the volume, respectively. In the case of the rutile phases, we depict the variations of the Ge-O-Ge (Si-O-Si) bond angles instead of the tilt angles since the rutile structure has a higher symmetry than the others. 
The models at elevated pressures do not exhibit the amorphous phase or transform into another phase because the calculated O-Ge-O and O-Si-O bond angles show no indication that the tetrahedra become significantly more irregular or distorted. Although the lattice constants are varied by increments of 1\%, the variations of the bond lengths are less than $\sim$0.1\%. This indicates that the bond angles play a predominant role in the compression or expansion of the fourfold oxides. Note that the bond lengths in the sixfold oxides change under pressure due to the higher symmetry of the O-Ge-O (O-Si-O) bond angles as well as the Ge-O-Ge (Si-O-Si) ones. 
The variations of the O-Ge-O bond angles are larger than those of O-Si-O [Figs.~\ref{fig:O_x_O}(a) and ~\ref{fig:O_x_O}(b)], whereas the tilt angles in SiO$_2$ vary more significantly than those in GeO$_2$ [Figs.~\ref{fig:x_O_x}(a) and ~\ref{fig:x_O_x}(b)] with respect to the volume. These results indicate the strong rigidity of the O-Si-O bonds. 
The experimental study reported that the O-Ge-O bond angles markedly change in q-GeO$_2$, while the tilt angle in q-SiO$_2$ varies significantly as the pressure increases, which agrees well with our result \cite{33}. The metallic property of Ge, as mentioned above, is also attributed to the distorted GeO$_4$ tetrahedra and the variation of the O-Ge-O bond angles from the ideal tetrahedral angle. This characteristic of the O-Ge-O bonds leads to markedly reduced lattice stress at the Ge/GeO$_2$ interface during the oxidation process compared with its Si counterpart. 
Kageshima and Shiraishi reported that SiO$_2$ forms a cristobalite structure at the initial stage of Si oxidation transforms into a quartz structure after the Si atom at the Si/SiO$_2$ interface has been ejected to release the interface stress \cite{kageshima}. We also investigated the emission of Ge atoms from the Ge/c-GeO$_2$ interface following their model and found that Ge atoms are hardly emitted \cite{saito}. We have concluded that this is because the dispersion of the bond angles around the suboxidized Ge atom at the Ge/GeO$_2$ interface is larger than that around the Si atom at the Si/SiO$_2$ interface. The present result that the bond angles around Ge atoms in c-GeO$_2$ also more drastically change than those around Si atoms as well as those in the other phases supports the conclusion in our preceding study.


\section{Conclusions}
We have calculated the bond lengths and bond angles of the rutile, $\alpha$-cristobalite, and $\alpha$-quartz phases of GeO$_2$ and SiO$_2$ using first-principles electronic-structure calculations. It was found that rutile GeO$_2$ is the most stable phase among the structures of GeO$_2$ examined here while SiO$_2$ preferentially forms the quartz structure, which agrees well with previous first-principles results. Symmetry allows four different O-Ge-O (O-Si-O) tetrahedral angles in GeO$_4$ (SiO$_4$) in the case of the cristobalite and quartz phases, and the angles in GeO$_4$ are more distorted from the ideal tetrahedral angle (109.5$^{\circ}$) than those in SiO$_4$ at the equilibrium volume. Moreover, we have also examined the variation of the bond lengths and bond angles with respect to volume and found that the mechanisms leading to compression and expansion are markedly different between GeO$_2$ and SiO$_2$ even though the volume compressibilities and expansibilities are almost identical: the tetrahedra of GeO$_4$ are significantly deformed under pressure whereas the tilting angle composed of two tetrahedras markedly varies in the case of SiO$_2$. These characteristics of GeO$_2$, i.e., the ground-state phase of the oxides and the difference in the variation of bond angles with respect to the volume, can be interpreted in terms of the metallic properties of the bond network of Ge. Thus, our results are highly relevant to the low defect density at the Ge/GeO$_2$ interface because the deterioration of the interface is suppressed owing to the flexibility of the O-Ge-O bond angles.

\acknowledgements
The authors would like to thank Professor Kikuji Hirose, Professor Yoshitada Morikawa, and Professor Heiji Watanabe of Osaka University, and Professor Kenji Shiraishi of University of Tsukuba for reading the manuscript and fruitful discussions. 
This research was partially supported by Strategic Japanese-German Cooperative Program from Japan Science and Technology Agency and Deutsche Forschungsgemeinschaft, by a Grant-in-Aid for Young Scientists B (No. 20710078), and also by a Grant-in-Aid for the Global COE "Center of Excellence for Atomically Controlled Fabrication Technology" from the Ministry of Education, Culture, Sports, Science and Technology, Japan. The numerical calculation was carried out using the computer facilities of the Institute for Solid State Physics at the University of Tokyo, Center for Computational Sciences at University of Tsukuba, the Research Center for Computational Science at the National Institute of Natural Science, and the Information Synergy Center at Tohoku University.

\begin{table*}
\caption{\label{tbl:lattice} Calculated and experimental lattice constants (unit: \AA).}
\begin{tabular}{lccccccccccccc} \hline
& \multicolumn{2}{c}{Present work} & \multicolumn{4}{c}{Other works} & \multicolumn{4}{c}{Experiment} \\
&$a$&$c$&&$a$&$c$&Ref.&&$a$&$c$&Ref. \\ \hline
q-GeO$_2$&4.897&5.636&&4.870&5.534&\citen{h}    &&4.987&5.652&\citen{smith} \\
c-GeO$_2$&4.818&7.128&&     &     &             &&4.985&7.070&\citen{boehm} \\
r-GeO$_2$&4.418&2.886&&4.283&2.782&\citen{h}    &&4.397&2.863&\citen{hazen} \\
q-SiO$_2$&4.850&5.348&&4.883&5.371&\citen{h}    &&4.916&5.405&\citen{25} \\
c-SiO$_2$&4.925&6.828&&4.950&6.909&\citen{h}    &&4.929&6.847&\citen{pluth} \\
r-SiO$_2$&4.147&2.662&&4.175&2.662&\citen{h}    &&4.180&2.667&\citen{51} \\
\hline
\end{tabular}
\label{tbl:1}
\end{table*}

\begin{table*}
\caption{\label{tbl:angle} Bond lengths $l_i$ (in \AA), bond angles $\theta_j$, and tilt angles $\delta$ (in deg) of GeO$_2$ and SiO$_2$ polymorphs. $\theta_1$, $\theta_2$, $\theta_3$, and $\theta_4$ are the O-x-O angles, where x represents a Ge or Si atom. $l_i$ and $\theta_j$ are assigned according to the magnitude. }
\begin{tabular}{llllllllllll}
\hline
&  & $l_1$ & $l_2$ & $\theta_1$ & $\theta_2$ & $\theta_3$ & $\theta_4$ & $\delta$ \\
\hline
q-GeO$_2$ & Present work&  1.763 & 1.755 & 114.13 & 110.69 & 107.28 & 105.39 & 29.66 \\
& Ref.~\citen{smith}  & 1.741 & 1.737 & 113.1  & 110.4  & 107.7  & 106.3  & 26.54 \\
c-GeO$_2$ & Present work&  1.760 & 1.760 & 120.69 & 111.39 & 109.95 & 101.72 & 35.64 \\
& --- \\
r-GeO$_2$ & Present work&  1.918 & 1.887 & 80.25 \\
& Ref.~\citen{hazen}   & 1.903 & 1.871 & 80.2  \\
q-SiO$_2$ & Present work&  1.608 & 1.603 & 110.58 & 109.37 & 109.23 & 108.55 & 17.85 \\
& Ref.~\citen{25}      & 1.614 & 1.605 & 110.52 & 109.24 & 108.93 & 108.81 & 16.37 \\ 
c-SiO$_2$ & Present work&  1.604 & 1.603 & 111.46 & 110.02 & 109.01 & 108.15 & 25.41 \\
& Ref.~\citen{32}      & 1.603 & 1.603 & 111.42 & 109.99 & 109.03 & 108.20 & 23.25 \\ 
r-SiO$_2$ & Present work&  1.786 & 1.751 & 81.02 \\
& Ref.~\citen{51}      & 1.810 & 1.758 & 81.35 \\ 
\hline
\end{tabular}
\end{table*}

\makefigurecaptions

\makefigurecaptions

\makefigurecaptions

\makefigurecaptions

\makefigurecaptions

\end{document}